\documentstyle[12pt,aaspp4,psfig]{article}

\begin{document}


\title{The Fastest Rotating Pulsar: A Strange Star?$^{*}$}

\author{XU Ren-Xin(~~~~~~~~~~), XU Xuan-Bin(~~~~~~~~~~), WU Xin-Ji(~~~~~~~~~~)\\
        CAS-PKU Joint Beijing Astrophysical Center
        and Department of Astronomy, \\
        Peking University, Beijing 100871}


\altaffiltext{0}{$^*$Supported by the National Science Foundation
of China under Grant No. 19803001, and the Special Funds for
Major State Basic Research Projects of China.}

\begin{abstract}

According to the observational limits on the radius and mass, the
fastest rotating pulsar (PSR 1937+21) is probably a strange star,
or at least some neutron star equations of state should be ruled
out, if we suggest that a dipole magnetic field is relevant to
its radio emission. We presume that the millisecond pulsar is a
strange star with {\em much} low mass, small radius, and weak
magnetic moment.

\vspace{0.4cm} %
\noindent %

\end{abstract}


Pulsars are conventionally modeled as neutron stars, but a strange
star$^{[\cite{hl97,cz01}]}$ model for pulsars is also proposed
(see also a review, e.g., by Xu et al.$^{[\cite{xzq01}]}$).
Therefore it is of great importance to find observationally
unambiguous features to distinguish strange stars from neutron
stars.
There are three hopeful ways known hitherto to identify a strange
star, which are based upon the differences of viscosity,
mass-radius relation, and surface condition between neutron and
strange stars.$^{[\cite{xu01}]}$
%
%
Recently, Kapoor and Shukre$^{[\cite{ks00}]}$ suggested to
constrain the equations of state of neutron stars by Rankin's
experiential line$^{[\cite{ran90}]}$ of the core emission of radio
pulsars, and find the restriction that the pulsar masses $M\leq
2.5M_\odot$ and radii $R\leq 10.5$ km, indicating that puslars
are strange stars.
However, Kapoor and Shukre's work$^{[\cite{ks00}]}$ has at least
two unseemly points: (1) the polar cap is defined for aligned
rotators (rather than for orthogonal rotators) in their
calculation; (2) the Rankin line is not doubtless true due to
many observational and statistical uncertainties. By redefining
the polar cap for orthogonal rotators, still we can {\em not}
obtain a conclusion that puslars are strange stars, or that some
neutron stars equations of state are ruled out, if the Rankin
line is used to constrain pulsar masses and radii. Nevertheless,
when applying this method to the fastest rotating pulsar, PSR
1937+21, we conclude that the pulsar may be a strange star, or at
least some equations of state should be ruled out.
In order to find solid evidence to shown that PSR 1937+21 is a
strange star, it is strongly suggested to measure precisely the
maximum rate of position-angle swing by further polarization
observations.

Now, we improve Kapoor and Shukre's computation by removing the
first unseemliness. In a spherical coordinate system with magnetic
axis $\mu$ being chosen as the z-axis, one has the following form
for a position vector {\bf r} in a dipole field line denoted by
parameter $\lambda$,$^{[\cite{af63}]}$
\begin{equation}
r=\lambda {cP\over 2\pi} \sin^2\theta,%
\label{r-theta}
\end{equation}
where $\theta$ is the polar angle, $P$ the rotation period, $c$
the speed of light, and $\lambda$ characterizes the sorts of
field lines.
Based on Eq.(\ref{r-theta}), the angle $\theta_\mu$ between ${\bf
\mu}$ and the direction of magnetic field at {\bf r} reads,
\begin{equation}
\cos\theta_\mu={2\lambda-3{\cal R}\over \sqrt{\lambda
(4\lambda-3{\cal R})}},%
\label{theta_mu1}
\end{equation}
where ${\cal R}\equiv {2\pi r\over cP} \ll 1$ for points near
pulsar surface. Expand function $\theta_\mu$ in the vicinity of
${\cal R }=0$, one comes to
\begin{equation}
\theta_\mu={3\over 2\sqrt{\lambda}}{\cal R}^{1/2}+{3\over
8\lambda\sqrt{\lambda}}{\cal R}^{3/2}+O({\cal R}^{5/2}).
\label{theta_mu}
\end{equation}
Assuming $\lambda=\lambda_0$ for the last-open-field lines (note:
$\lambda_0$ is a fuction of inclination angle $\alpha$), one can
obtain the beam radius $\rho$ for the radiation at an emission
height $h=r-R$ ($R$ is the radius of pulsar), neglecting the
terms being equal or higher than ${\cal R}^{3/2}$ in
Eq.(\ref{theta_mu}),
\begin{equation}
\rho={3\over 2}\sqrt{2\pi\over \lambda_0 c} {r^{1/2}\over
P^{1/2}}={1.24^{\rm o}\over \sqrt{\lambda_0 P} }\sqrt{r\over
10{\rm km}}.
\label{rho1} %
\end{equation}

For core emissions ($r\ga R$), which are supposed to be originated
from regions near pulsar surfaces,$^{[\cite{ran90}]}$ general
relativistic effects are not negligible due to the spacetime
curvature.$^{[\cite{ks98}]}$ Two such effects, squeezing of the
dipole magnetic field and bending of the radio wave, can be
represented approximately by two factors,$^{[\cite{ks98,ks00}]}$
$f_{\rm sqz}$ and $f_{\rm bnd}$, respectively, so that the beam
radius $\rho$ can be re-written as
\begin{equation}
\rho={3\over 2}\sqrt{2\pi\over \lambda_0 c} {r^{1/2}\over
P^{1/2}}={1.24^{\rm o}\over \sqrt{\lambda_0 P} }\sqrt{r\over
10{\rm km}}f_{\rm sqz} f_{\rm bnd},
\label{rho} %
\end{equation}
where $f_{\rm sqz}=(1+{3GM\over 2c^2r})^{-1/2}$, $f_{\rm
bnd}={1\over 3}(2+(1-{2GM\over c^2r})^{-1/2})$, $G$ is the
gravitation constant, $M$ the pulsar mass.

We are to calculate the parameter $\lambda_0$ of last-open-field
lines below, assuming that only the plasma within the light
cylinder with radius $r_{\rm lc}={cP/(2\pi)}$ can corotates with
the star, i.e., field lines that penetrate beyond the light
cylinder should be open. Generally, $\lambda_0$ is a function of
inclination angle $\alpha$. Nevertheless, one can obtain simple
formulae of $\lambda_0$ in two special cases. Obviously,
$\lambda_0(\alpha=0)=1$ for an aligned rotator according to
Eq.(\ref{r-theta}).
Another formula of $\lambda_0$ can be easily reached for an
orthogonal rotator ($\alpha=90^{\rm o}$). The conal angle
$\theta_0$ of the null surface, where the magnetic fields are
perpendicular to magnetic axis $\mu$, can be obtained from
Eq.(\ref{theta_mu1}) by setting $\theta_\mu=90^{\rm o}$,
\begin{equation}
\theta_0=\sin^{-1}\sqrt{{2\over 3}}.%
\label{theta0} %
\end{equation}
Let $r\cos\theta_0=r_{\rm lc}$, one finds
\begin{equation}
\lambda_0(\alpha=90^{\rm o})={3\sqrt{3}\over 2},%
\label{lambda0} %
\end{equation}
which is 2.6 times of $\lambda_0(\alpha=0^{\rm o})$.

\vspace{0.0cm}%
\centerline{} \centerline{%
\psfig{file=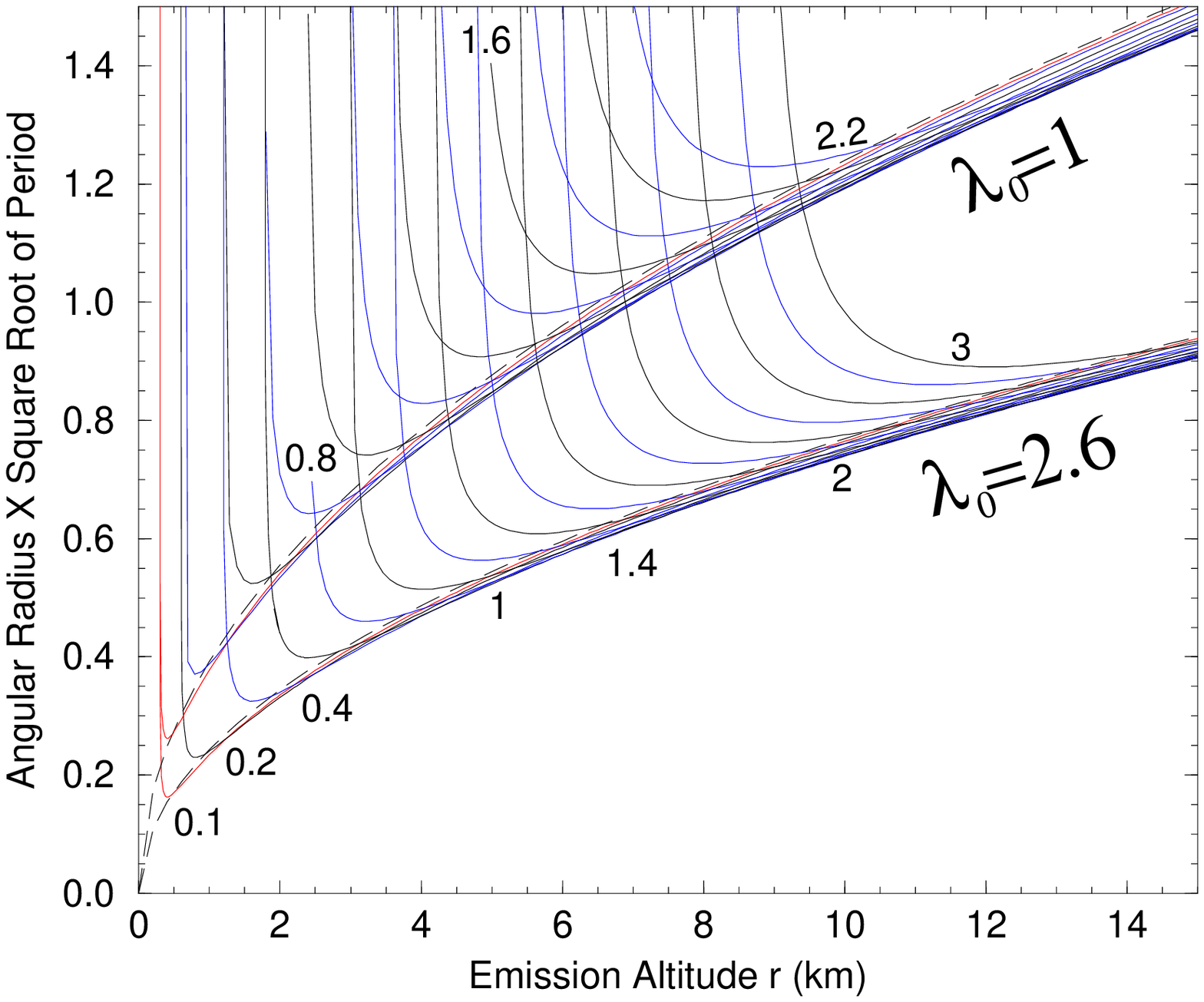,width=10cm,height=8cm}} %
\figcaption{Calculated results based on Eq.(\ref{rho}) show the
variations of $\rho\times\sqrt{P}$ (in deg$\cdot$s$^{1/2}$) as a
function of emission altitude $r$ for different pulsar masses
ranging from $0.1M_\odot$ to $\ga 2M_\odot$. We present results
of $\lambda_0=1$ (aligned rotators) and $\lambda_0=3\sqrt{3}/2$
(orthogonal rotators) for two limits; an actual value for a
pulsar with artificial inclination angle $\alpha$ should be
between those two corresponding values. The dashed lines are
calculated by Eq.(\ref{rho1}), which are not relevant to pulsar
mass.
\label{f.1}}%
\vspace{0.2cm}

Following the representation by Kapoor and
Shukre,$^{[\cite{ks00}]}$ we calculate the value,
$\rho\times\sqrt{P}$, as a function of emission altitude $r$ for
pulsars with core emission according to Eq.(\ref{rho}). The
calculated results are shown in Fig.\ref{f.1}. Contrary to the
result of Kapoor and Shukre,$^{[\cite{ks00}]}$ it is clear from
Fig.\ref{f.1} that the Rankin line (i.e.,
$\rho\times\sqrt{P}=1.225^{\rm o}~{\rm s}^{1/2}$) can not
effectively put any limit on the equations of state of neutron
stars or conduce toward any indication that pulsars are strange
stars if we assume that Rankin's samples are of nearly orthogonal
rotators. Nevertheless, we may still obtain some valuable
conclusions from above calculation for some special pulsars, for
instance, the fastest rotating millisecond pulsar, PSR 1937+21.

The rotation period of PSR 1937+21 is $P=1.558$ ms, the smallest
one of pulsars observed, and the rate of period change is $\dot
P=10^{-19}~{\rm s\cdot s}^{-1}$. The radius of light cylinder of
the star is $r_{\rm lc}=cP/(2\pi)=74$km . The corotation radius,
defined by the balance of gravitational force and the centrifugal
force, is $r_{\rm c}=[GM/
(4\pi^2)]^{1/3}P^{2/3}=20(M/M_\odot)^{1/3}$km, where $M$ is the
mass of PSR 1937+21. Therefore the pulsar radius $R$ satisfies
\begin{equation}
R<{\rm min}(r_{\rm lc},r_{\rm c})=20(M/M_\odot)^{1/3}\;{\rm km}.%
\label{R<rc}
\end{equation}

Another circumscription of the stellar radius $R$ and the mass
$M$ may arise from the inclusion of general relativistic
effect.$^{[\cite{ks00}]}$ The pulse width $\Delta \phi$ of PSR
1937+21 is about $10^{\rm o}$ at 600 MHz and 1.4 GHz (EPN
database), which is difficult to be explained geometrically if we
suggest PSR 1937+21 has a radius of 10 km for canonical
pulsars.$^{[\cite{cs83}]}$ Gil propose thus a
model$^{[\cite{g85}]}$ in which a quadrupole magnetic field
geometry is assumed. However an alternative conjecture to
overcome the difficulty is that PSR 1937+21 may have {\em
unusual} small radius. Taking the simplest proposing that
$\alpha=90^{\rm o}$ and impact angle $\beta=0$, we know
$\Delta\phi=2\rho$, and therefore $\rho\sqrt{P}=0.2^{\rm o}~{\rm
s }^{1/2}$. According to the results in Fig.\ref{f.1}, it is
found that
\begin{equation}
M<0.2~M_\odot~~{\rm and} ~~R<1 ~{\rm km.}
\label{limit0}
\end{equation}
These {\em stringent } limits have to result in the conclusion
that PSR 1937+21 is a {\em strange star} rather than a neutron
star because of the strikingly different mass-radius relations.

However, the conclusion in Eq.(\ref{limit0}) is {\em not} so
solid. In fact there are two models proposed to account for the
inter-pulse emission:$^{[\cite{g83}]}$ single-pole model and
double-pole model. Three observational facts$^{[\cite{g85}]}$
favour a double-pole model of inter-pulse origin of PSR 1937+21.
(1) The gradient of the position angle at inter-pulse center has
the same value as at the main-pulse center. (2) The separation of
the main-pulse and the inter-pulse is nearly 180$^{\rm o}$. (3)
The intensities of the main-pulse and the inter-pulse are roughly
equal. The observed maximum rate of position angle swing at the
main-pulse center is $({\rm d}\psi / {\rm d}\phi)_{\rm
max}\approx 3$. In the standard rotating vector model, the impact
angle $\beta$ should be
\begin{equation}
\beta =\sin^{-1}[{\sin\alpha \over ({\rm d}\psi / {\rm
d}\phi)_{\rm max}}]
\sim 19.5^{\rm o},%
\label{beta}
\end{equation}
and the beam radius $\rho$ thus can be derived as
\begin{equation}
\rho=\cos^{-1}[\cos\beta-2\sin\alpha\sin(\alpha+\beta)\sin^2
{\Delta\phi\over 4}]\sim 20.1^{\rm o},%
\label{rho}
\end{equation}
if we take $\alpha=90^{\rm o}$, $({\rm d}\psi / {\rm d}\phi)_{\rm
max}=3$ and $\Delta\phi=10^{\rm o}$. In this case, we find
$\rho\sqrt{P}=0.79^{\rm o}~{\rm s }^{1/2}$, and the limits thus
are
\begin{equation}
M<2.4~M_\odot~~{\rm and} ~~R<11.5 ~{\rm km}
\label{limit}
\end{equation}
based on Fig.\ref{f.1}.

\begin{deluxetable}{lcr}
\tablewidth{5.0in} \tablenum{1}%
\tablecaption{Total gravitational mass-energy $M$ and the
circumferential Radii $R$ at equator of rotating ($P=1.558~{\rm
ms}$) neutron stars with maximum mass for five kinds of
Equations of state}%
\tablehead{ \colhead{EOS$^*$} & \colhead{$M/M_\odot$} &
\colhead{$R$(km)} }
\startdata%
A (Reid soft core) & 1.6604 & 8.82 \nl%
L (Mean field) & 2.7263 & 14.98 \nl%
M (Tensor interaction) & 1.8298 & 19.08 \nl%
AU (AV14 + UVII) & 2.1433 & 9.78 \nl%
FPS (UV14 + TNI) & 1.8069 & 9.88 \nl%
\enddata
{$^*$ See Table 2 of Cook et al.$^{[\cite{cst94}]}$ for details.}%
\vspace{0.2cm}
\end{deluxetable}{}

It is well known that the masses and radii of rotating magnetized
neutron stars are of equation-of-state dependent. The radius of a
neutron star with a maximum mass is the smallest; a smaller mass,
a slower rotation and/or a lower inner magnetic field would cause
its radius to be larger.$^{[\cite{cst94} - \cite{bbgn95}]}$
Therefore the observational deduction of Eq.(\ref{limit}) should
have some implications for proof-testing the equations of state
available. Five euaqtions of state have been
focused$^{[\cite{cst94}]}$ in the calculation for rapidly
rotating neutron stars, the results of which are listed in Table
1 by interpolating between the tabulated points in Table 9-23 of
Cook et al.$^{[\cite{cst94}]}$ for angular frequency
$\Omega=4\times 10^3~{\rm s}^{-1}$ (i.e., $P=1.558$ ms). We see
that the equations labeled as ``L'' and ``M'' should be ruled out
by the limits of Eq.(\ref{limit}).
Certainly, we should also keep in mind that more equations of
state may be killed by inclusion of the inner magnetic field
effect,$^{[\cite{bbgn95}]}$ which is an interesting topic in the
future study.

Furthermore, two observational uncertainties may result in a
derivation of stronger limits on the mass and radius of PSR
1937+21. The first is that observation would give a less steep
position angle gradient (i.e., a smaller value of $({\rm d}\psi /
{\rm d}\phi)_{\rm max}$ than reality) due to smearing of finite
sampling time, to the frequency dispersion in pulse arrive
time,$^{[\cite{lw99}]}$ and to the interstellar
scattering$^{[\cite{g85}]}$ which may bring on a stretching out of
the longitude scale. In fact, $({\rm d}\psi / {\rm d}\phi)_{\rm
max}>3$. A larger $({\rm d}\psi / {\rm d}\phi)_{\rm max}$ (thus a
smaller $\beta$ according to Eq.(\ref{beta})) favours a rough
equality of the intensities of the main-pulse and inter-pulse.
The second is that PSR 1937+21 might not be exactly an orthogonal
rotator (i.e., $\alpha$ is not precisely $90^{\rm o}$). It is
possible that $\alpha<90^{\rm o}$.
Both these two uncertainties lead to a smaller value of
$\rho\sqrt{P}$. We thus take $({\rm d}\psi / {\rm d}\phi)_{\rm
max}$ and $\alpha$ as two free parameters to calculate
$\rho\sqrt{P}$. The calculated results are shown in Fig.\ref{f.2}.
It is clear that stronger limits than those of Eq.(\ref{limit})
may rule out more neutron star equations of state, and probably
conduce toward a strange star model for PSR 1937+21, based on
Figs.\ref{f.1} and \ref{f.2}.

\vspace{0.0cm}%
\centerline{} \centerline{%
\psfig{file=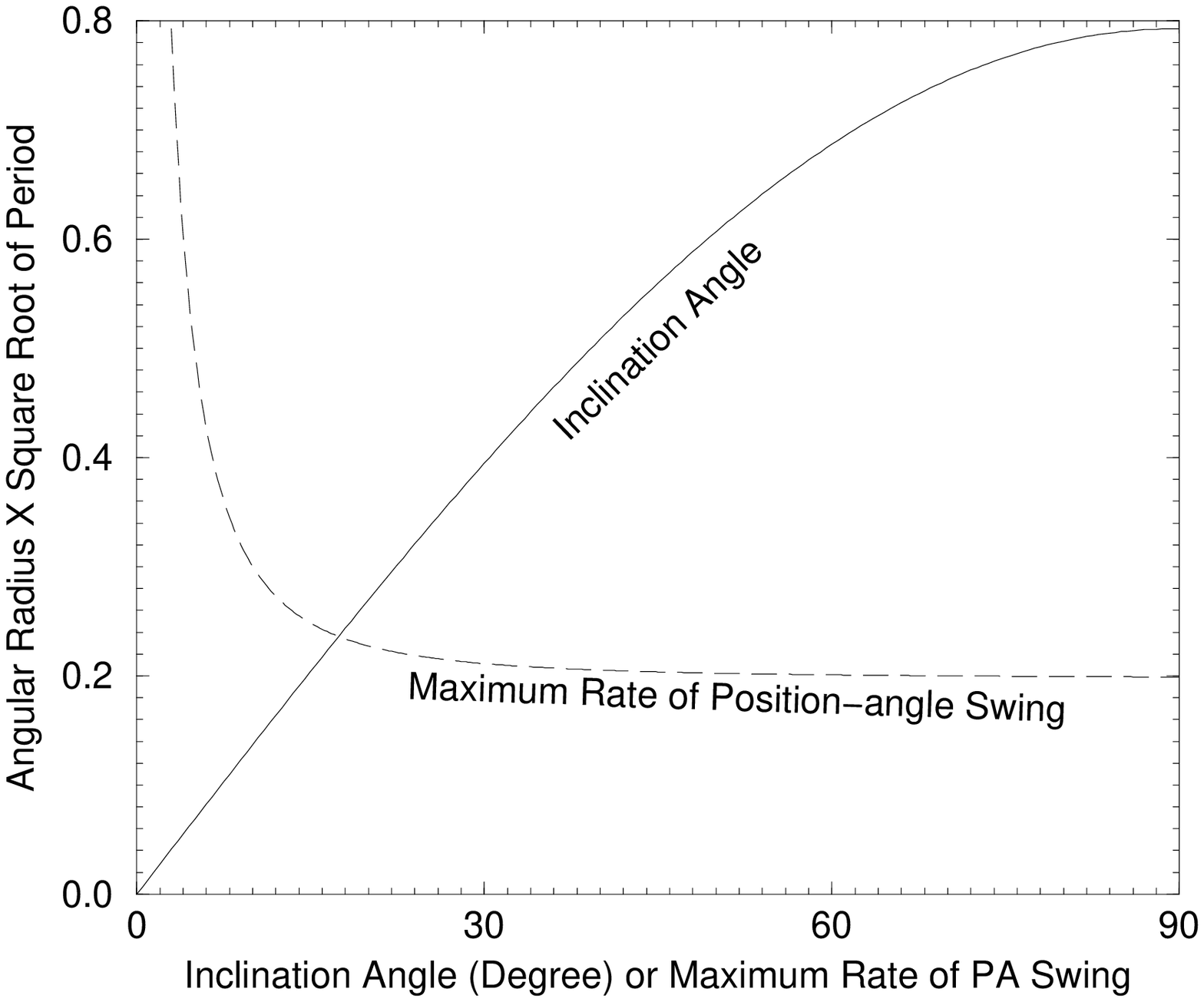,width=10cm,height=8cm}} %
\figcaption{Beam angular radius times square root of period
($\rho\sqrt{P}$) of PSR 1937+21 versus possible different
inclination angles $\alpha$ for the case of $({\rm d}\psi / {\rm
d}\phi)_{\rm max}=3$, or maximum rates of position-angle swing
$({\rm d}\psi / {\rm d}\phi)_{\rm max}$ for $\alpha=90^{\rm o}$.
\label{f.2}}%
\vspace{0.2cm}

In Fig.\ref{f.2} we find $\rho\sqrt{P}$ is sensitively dependent
on $({\rm d}\psi / {\rm d}\phi)_{\rm max}$ if $({\rm d}\psi /
{\rm d}\phi)_{\rm max}\la 10$. $\rho\sqrt{P}=0.6$ if $({\rm
d}\psi / {\rm d}\phi)_{\rm max}=4$ and $\alpha=90^{\rm o}$, in
this case the limits of Eq.(\ref{limit}) should be modified as
$M<1.4~M_\odot$ and $R<6.6$ km. These limits inevitably lead to
the conclusion that PSR 1937+21 is a strange star. Because of the
observational uncertainties of $({\rm d}\psi / {\rm d}\phi)_{\rm
max}$ and $\alpha$ and of the importance of distinguishing a
strange star in nature, we seriously table a proposal of
detecting more accurate polarization signal from PSR 1937+21 to
obtain an actual value, especially of $({\rm d}\psi / {\rm
d}\phi)_{\rm max}$. This could be possibly by means of reducing
the sampling time and increasing the total observation time.

If PSR 1937+21 is a strange star with small radiu ($R<10$ km),
the old formulae$^{[\cite{mt77}]}$ to calculate the surface
magnetic field $B$ and dipole magnetic moment $\mu_{\rm m}$
should be modified as
\begin{equation}
\begin{array}{lll}
B & = & ({c^3 \rho P \dot P \over 5\pi R})^{1/2}
\sim 9.3\times 10^{19} R_5^{-1/2} (P \dot P)^{1/2}~~{\rm G},\\
\mu_{\rm m} & = & ({c^3 \rho P \dot P R^5 \over 20\pi})^{1/2}
\sim 4.6\times 10^{29} R_5^{5/2} (P \dot P)^{1/2} ~~{\rm G\cdot
cm^3},
\end{array}
\label{B}
\end{equation}
since the density $\rho\sim 5\times 10^{15}$ g cm$^{-3}$ has a
very modest variation with radial distance of strange
star,$^{[\cite{afo86}]}$ $M\sim (4/3)\pi R^3\rho$, and the moment
of inertia $I\sim (8/15) \pi \rho R^5$, where $R_5$ is the stellar
radius in $10^5$ cm (1 km). We see from Eq.({\ref{B}}) that the
calculated magnetic moment is strongly related to the radius
although the observation-determined surface magnetic field is
weakly related to. For instance, if the radius of PSR 1937+21 is
only 1 km, $B=1.2\times 10^9$ G, whilst $\mu_{\rm m}=5.8 \times
10^{18}$ G$\cdot$cm$^3$ (typically, $\mu_{\rm m}$ for millisecond
pulsars is assumed to be $10^{26}$ G$\cdot$cm$^3$). This property
favors the assumption that dipole magnetic structure dominates in
the radio emission region.


Millisecond pulsars are currently believed to be of
recycled-origin of normal pulsars which are spin down enough.
However, motivated by the study of planet formation around PSR
1257+12, Miller and Hamilton$^{[\cite{mh01}]}$ suggested that some
and perhaps all isolated millisecond pulsars may have been born
with high spin rates and low magnetic fields instead of having
been recycled by accretion. This is understandable if we assume
part or all of the isolated millisecond pulsars are strange stars
with smaller radii (thus smaller masses): (1) Turbulent convection
in nuclear matter$^{[\cite{td93}]}$ or in strange quark
matter$^{[\cite{xb01}]}$ should be less prosperous in a
proto-pulsar with lower mass than with higher mass in the
Kelvin-Hermholtz cooling phase, the dynamo-created magnetic field
is thus weaker. (2) Small radius favors faster rotation since the
centrifugal force become smaller while the gravitational force to
be larger (see Eq.(\ref{R<rc})). A bimodal distribution of pulsar
periods and magnetic fields may arises from various kind of
progenitors as well as complex mechanisms of supernova
exploration.
Therefore, it is conjectured that PSR 1937+21 and PSR 1257+12 (and
possibly some or all of the isolated millisecond pulsars) may
have weaker dynamo-originated magnetic field due to a less
effective field magnification process.

\vspace{1cm}

\noindent %

\end{document}